\documentclass[10pt,final,journal]{IEEEtran}

\usepackage{amsmath}
\usepackage{amsfonts}
\usepackage{amssymb}
\usepackage{graphicx}

\begin{document}

\title{The Internet's Unexploited Path Diversity}

\author{
	\IEEEauthorblockN{P. D. Arjona-Villica\~{n}a, C. C. Constantinou and A. S. Stepanenko}
	\thanks{P. D. Arjona-Villica\~{n}a and C. C. Constantinou are with the University of Birmingham, UK (e-mail:pda574@bham.ac.uk).}
	\thanks{A. S. Stepanenko is with Aston University, Birmingham, UK.}
	\thanks{This work was supported by CONACYT (Mexico).}
}
\maketitle

\begin{abstract}
The connectivity of the Internet at the Autonomous System level is influenced by the network operator policies implemented. These in turn impose a direction to the announcement of address advertisements and, consequently, to the paths that can be used to reach back such destinations. We propose to use directed graphs to properly represent how destinations propagate through the Internet and the number of arc-disjoint paths to quantify this network's path diversity. Moreover, in order to understand the effects that policies have on the connectivity of the Internet, numerical analyses of the resulting directed graphs were conducted. Results demonstrate that, even after policies have been applied, there is still path diversity which the Border Gateway Protocol cannot currently exploit.
\end{abstract}


\section{Introduction}\label{intro}

The \emph{Border Gateway Protocol} (BGP) is currently the only routing protocol connecting the Internet at the Autonomous System (AS) level \cite{Rekhter2006}. Ever since its deployment, BGP has been criticized because it is prone to  stability and convergence problems \cite{Labovitz1997, Labovitz2000}. Before any significant improvements to BGP can be implemented, we advocate that we need a better fundamental understanding of the impact of policies in shaping the Internet's effective, or available, topology.

Previous studies of the Internet's topology \cite{Mahadevan2006} have focused on obtaining an undirected graph $G=\left(V,E\right)$, in which vertices, $V$, represent ASs and edges, $E$, describe the bidirectional communication links that join them, and then examining the topological characteristics of this graph using different metrics. In reality, Internet service providers (ISP) are not only interested in obtaining a certain degree of connectivity; they also need to restrict access to their network by other competing ASs, while maintaining global end-to-end connectivity for all their host addresses. This gives rise to a contradiction in the objectives of BGP in that it is responsible for linking ASs, while at the same time diminishes the number of paths available in this network. In this letter we want to analyze the connectivity properties of the Internet which are allowed by the ISPs, regardless of the non-policy related restrictions imposed by the BGP routing protocol.

Routers may know and use more than one path to reach a destination. This path diversity is very useful to implement traffic engineering, or to define backup paths that can be used to avoid congestion and cope with link failures. However, we argue that BGP was not originally designed to fully exploit the path diversity available in the Internet.

This is because first, policies which allow ASs to decide which paths to accept, employ and propagate, are mostly used to maintain a hierarchy based on business agreements \cite{Gao2000a}. Although policies limit the path diversity available in the Internet, there is no fundamental reason why we cannot still use the paths that have survived after policies have been applied and destinations have been allowed to propagate.

Second, BGP only allows employing one preferred path to reach each destination in its routing table. We call this the \textbf{Preferred Paths Rule} (PPR). Originally, the PPR meant that BGP would pick the best route according to its own selection algorithm; in practice, it may cause a single path to become overwhelmed by the amount of traffic it needs to carry. Although there are mechanisms to overcome the PPR, these inherently contradict BGP's original functionality.

The path diversity of the Internet diminishes to accommodate the restrictions imposed by policies and the PPR. But how many more paths does an AS know when they ignore the PPR? Could the use of these paths provide any advantages over the current functionality provided by BGP? These are the questions we try to address in this letter.

\section{How Policies Modify the Internet's Topology}\label{how_policy_mod_topology}

\begin{figure}[!t]
	\centering
		\includegraphics[width=2.7in]{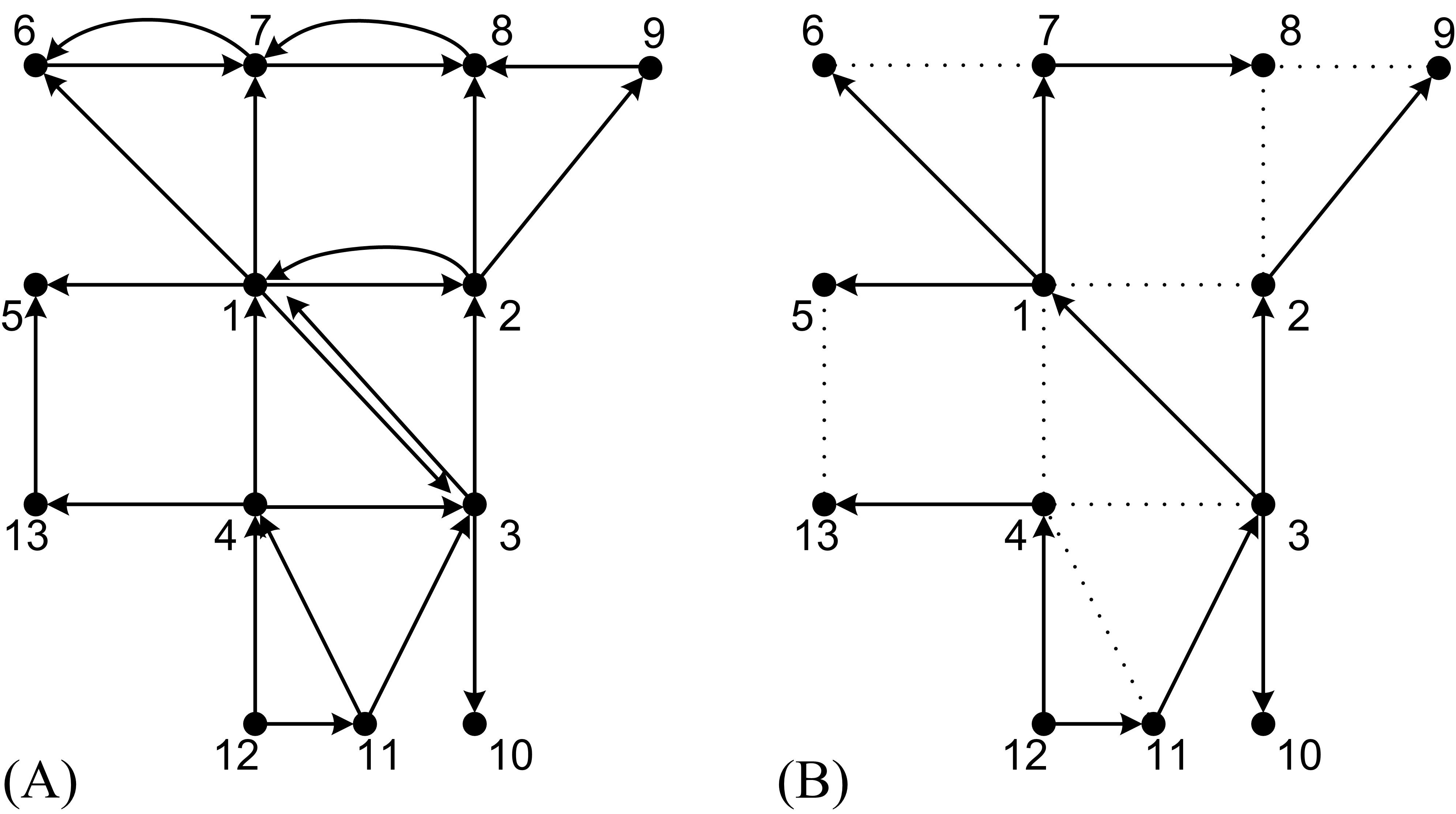}
	\caption{Announcement digraph (A) and BGP digraph (B) for AS12}
	\label{fig:42_AnnDigraph}
\end{figure}

Since ASs need to manage how to communicate destinations to their neighbors, policy filtering is an essential feature of any inter-AS routing protocol. It is thus necessary to represent the propagation of destinations as a \emph{directed graph} or \emph{digraph} $D=\left(V,A\right)$ which is just an orientation of $G$ (i.e., $A$ represents directed edges or arcs). The graph we use to represent how autonomous system $i$, AS\textit{i}, propagates destination advertisements through $G$ is called the \textbf{announcement digraph of AS\textit{i}}, $D_{a}\left(i\right)$. This graph is only restricted by the policies that ASs impose over their neighbors for transmitting the reachability of AS\textit{i}. Therefore, the announcement digraph may be considered as a union of all the permitted alternatives to transmit a destination through the Internet before the full BGP functionality is imposed. Fig.~\ref{fig:42_AnnDigraph}.A shows the announcement digraph $D_{a}\left(12\right)$.

As each AS selects its preferred route to reach a destination, alternative routes in the announcement digraph will be ignored in order to comply with BGP's PPR. Each AS will choose which routes to accept and which routes to announce depending on the policies defined by the network administrator. Fig.~\ref{fig:42_AnnDigraph}.B is an example of how the PPR modifies the announcement digraph for a destination and creates rigid paths which form an \emph{oriented tree} or \emph{arborescence}, which we call the \textbf{BGP digraph of AS\textit{i}}, $D_{BGP}\left(i\right)$. This graph is equivalent to what other authors have called a \emph{solution} to a BGP system \cite{Griffin1999}, because when this arborescence is defined, all the ASs in the network converge to the same solution. Because the BGP digraph provides only one path between AS\textit{i} and each other AS, when an arc (link) is lost the ASs that cannot reach back to AS\textit{i} need to find a new alternative path by consulting their routing tables and selecting a new path. This period in which ASs are processing and finding a new solution is what causes transient instabilities in a network.

In a digraph, the operation of \emph{reversing} an arc means to replace an arc $A = ab$ with an arc $A^{\prime} = ba$ that has the opposite direction. When we reverse all the arcs in a digraph we obtain the \emph{converse} of the same digraph. In this study, the converse of $D_{a}\left(i\right)$ represents the paths that other ASs may use to reach the destination originally advertised. We call this the \textbf{destination digraph of AS\textit{i}}, $D_{d}\left(i\right)$.

By comparing the announcement digraph and the BGP digraph (Fig.~\ref{fig:42_AnnDigraph}), it is evident that the former has more paths than the later, but still there is no clear way to measure this diversity. The literature on digraphs \cite{Bang-Jensen2002} provides two different definitions to quantify path diversity: A group of \emph{arc-disjoint paths} is a set of paths connecting two vertices through intermediate vertices, in which none of the paths traverse the same arc more than once; and a group of \emph{internally disjoint paths} is a set of paths connecting two vertices where neither the arcs nor the vertices are common in any of these paths. Because vertices in these digraphs represent ASs, which have a significant less probability of failure than their communication links or arcs, we recommend the number of \textbf{arc-disjoint paths, $adp$}, as the more meaningful unit for measuring the path diversity of an inter-AS network.

\section{The Internet's Path Diversity}\label{internet_path_div}

To obtain announcement digraphs that are representative of the path diversity that the Internet could potentially enjoy, we carried out a numerical experiment restricted to ASs with high connectivity. Since one of our main sources of data only provides routing information for Europe, this analysis' sample had to be restricted to European ASs. Also ASs closer to the core were selected because the ones at the edge of the Internet tend not to have enough connectivity, with a few exceptions \cite{Oliveira2008}. Therefore, this experiment was limited to the 44 largest European ASs, plus a global AS which supplied its routing information (AS3356). Henceforth, this set of core ASs will be termed the \textbf{Top-45} and they are listed in Table \ref{tab:TheTop45}. The procedure used to select these ASs is further described in \cite{Arjona-Villicana2009}. Routing data for the Top-45 was collected from the following sources:

\textbf{RIPE Database} \cite{RIPEDatabase} is a repository of routing policies provided by mostly European ASs. The information stored in this database allows to build topologies which represent how destinations could propagate if only the policies declared by each AS are implemented.

\textbf{RouteViews} \cite{Routeviews} and \textbf{RIPE RIS} \cite{RIPERIS}, are similar projects that collect snapshots of BGP routing tables. Paths stored at these tables enable to construct a topological map of how destinations propagate through the Internet.

\textbf{Skitter} \cite{Skitterproject} uses Internet Control Message Protocol (ICMP) to record the ASs visited when this message propagates through the Internet. Skitter is able to detect the adjacency between two ASs, but it cannot reveal the policies that apply between them, nor where a destination originated.

Data from these sources was harvested in two different dates to avoid the possibility of taking samples on a non-typical day: November 29th and December 19th, 2007. Since the results in both days are similar, here we only report the ones obtained in November 29th. Complete results can be consulted at \cite{Arjona-Villicana2009}.

Perl scripts were developed to parse and process the data from each of these data sources. The data needed to be converted into a format that could be easily analyzed and displayed; therefore adjacency matrices were selected. The \emph{adjacency matrix} of a digraph is a square matrix of dimensions $V \times V$ (the order or number of vertices in the digraph) in which each column (row) of the matrix represents the head (tail) of an arc. Each element of this matrix is either 1, if the arc that corresponds to the row-column pair is present in the digraph, or 0 otherwise. Since the data sources represent different views from the same network, it was concluded that in order to obtain the closest aproximation to the Internet's announcement digraphs it was necessary to combine their information by performing a union of the topology from all four sources.

\begin{figure}[!t]
		\includegraphics[width=3.2in]{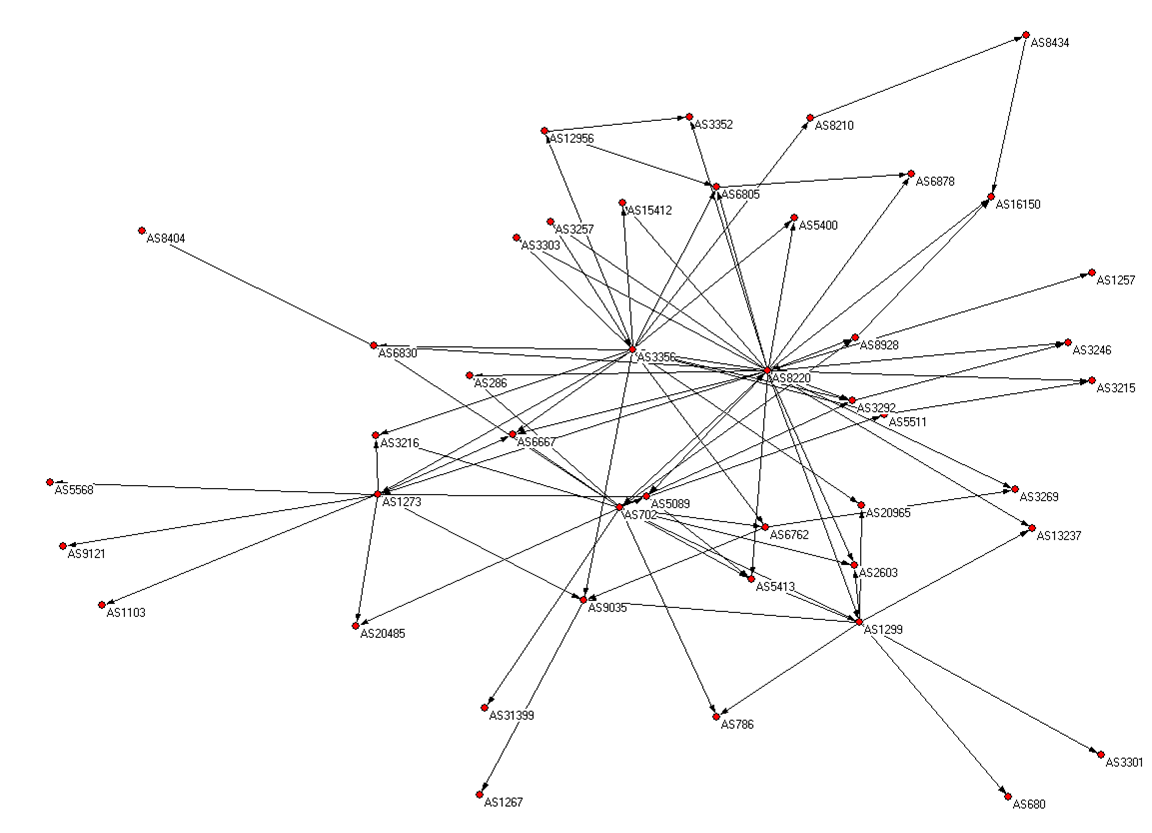}
	\caption{Announcement digraph obtained for AS8220}
	\label{fig:AnncDigraph}
\end{figure}

The 45 announcement digraphs $D_{a}\left(i\right)$ obtained represent how destination advertisements could propagate from an originating AS\textit{i}, to the other 44 ASs in the European core. Most of them were connected digraphs where AS\textit{i} is the origin. Many contained cycles and, in some, it was possible to return to AS\textit{i}. Just 9 digraphs had a few ASs that could not be connected from AS\textit{i}. We surmise that these ASs may be obtaining announcements through ASs not included in the Top-45 list. Fig.~\ref{fig:AnncDigraph} shows AS8220's announcement digraph. The adjacency matrices and announcement digraphs for the other Top-45, for both collection dates, can be downloaded from \cite{Arjona-Villicana2009}.

\begin{table}[!t]
  \renewcommand{\arraystretch}{1.2}
	\caption{Number of Arc-disjoint Paths between the Top-45 ASs}
	\label{tab:TheTop45}
	\centering
		\begin{tabular}{|c|c c c||c|c c c|}
			\hline	\textbf{AS} & \textbf{Avg} & \textbf{Min} & \textbf{Max} &
							\textbf{AS} & \textbf{Avg} & \textbf{Min} & \textbf{Max} \\
      \hline
      \hline	1299	&	1.61	&	1	&	4	&	3301	&	1.11	&	 1	&	2	\\
      \hline	702		&	2.59	&	1	&	6	&	8434	&	1.82	&	 1	&	4\\
      \hline	3303	&	2.77	&	1	&	5	&	5089	&	3.20	&	 1	&	6\\
      \hline	1257	&	1.77	&	0	&	4	&	6878	&	1.75	&	 1	&	2\\
      \hline	13237	&	2.20	&	1	&	4	&	31399	&	2.39	&	 1	&	3\\
      \hline	8220	&	2.00	&	1	&	4	&	6667	&	2.11	&	 1	&	4\\
      \hline	286		&	1.75	&	0	&	3	&	1103	&	1.89	&	 0	&	4\\
      \hline	3257	&	2.30	&	1	&	5	&	2603	&	1.93	&	 1	&	4\\
      \hline	1273	&	2.05	&	1	&	4	&	680		&	2.09	&	 1	&	4\\
      \hline	16150	&	2.59	&	1	&	5	&	3269	&	1.30	&	 0	&	2\\
      \hline	8928	&	2.36	&	1	&	4	&	3215	&	1.00	&	 0	&	2\\
      \hline	8342	&	2.09	&	1	&	4	&	9121	&	3.55	&	 1	&	5\\
      \hline	5413	&	1.95	&	1	&	4	&	786		&	2.36	&	 1	&	4\\
      \hline	5511	&	1.70	&	1	&	4	&	8404	&	1.00	&	 1	&	1\\
      \hline	12956	&	2.11	&	1	&	5	&	20485	&	3.73	&	 1	&	6\\
      \hline	6762	&	1.55	&	0	&	4	&	9035	&	2.18	&	 0	&	4\\
      \hline	15412	&	2.09	&	1	&	3	&	1267	&	1.77	&	 1	&	3\\
      \hline	5400	&	1.77	&	1	&	4	&	3352	&	1.07	&	 1	&	2\\
      \hline	20965	&	1.52	&	1	&	2	&	6830	&	2.59	&	 1	&	5\\
      \hline	3292	&	2.43	&	1	&	5	&	3216	&	3.80	&	 1	&	5\\
      \hline	3246	&	2.09	&	1	&	4	&	5568	&	3.66	&	 1	&	6\\
      \hline	6805	&	1.86	&	1	&	3	&	3356	&	1.80	&	 0	&	4\\
      \hline	8210	&	1.11	&	0	&	3	&	&	&	&	\\
			\hline
		\end{tabular}
\end{table}

As described in section \ref{how_policy_mod_topology}, the converse of an announcement digraph $D_{a}\left(i\right)$ can be used to obtain the corresponding destination digraph $D_{d}\left(i\right)$, which then allows to determine how many arc-disjoint paths, $adp$, each AS could use to reach AS\textit{i}. The result of this analysis for each of the Top-45 ASs produces a  large ($45 \times 45$) table (available in \cite{Arjona-Villicana2009}). More important is the number of arc-disjoint paths each AS can potentially use to reach the other Top-45 ASs. Table \ref{tab:TheTop45} shows the average, minimum and maximum number of $adp$ between each AS and the other 44 ASs. ASs that have 0 as the minimum illustrate the few ASs that could not be connected from AS\textit{i}. The frequency of each $adp$ value, in the $45 \times 45$ table \cite{Arjona-Villicana2009}, produces the histogram of Fig.~\ref{fig:num_arc_disj_paths}.

\section{Discussion and Conclusion}\label{discussion_and_conlcusion}

\begin{figure}[!t]
	\centering
		\includegraphics[width=2.5in]{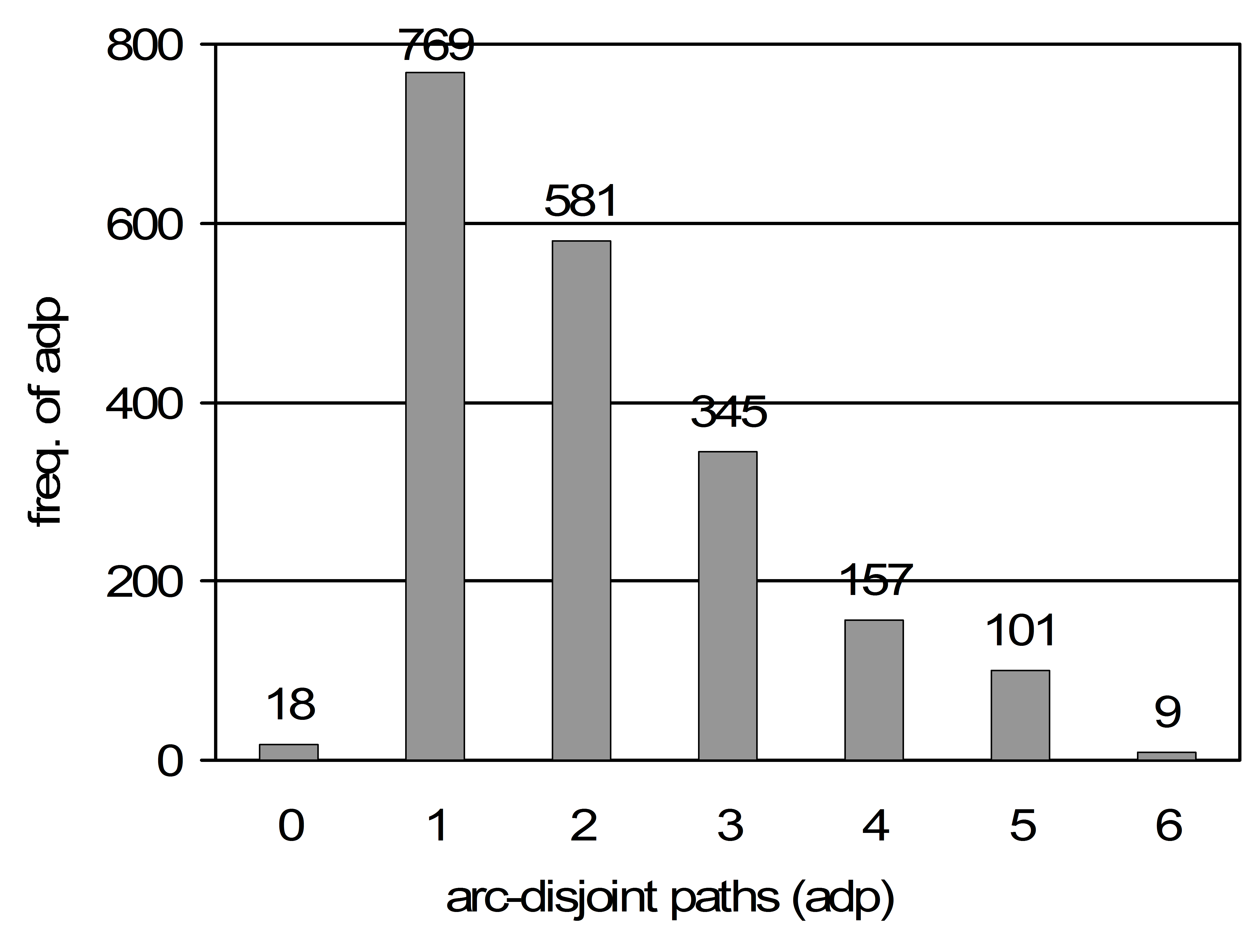}
	\caption{Number of arc-disjoint paths found}
	\label{fig:num_arc_disj_paths}
\end{figure}

Fig.~\ref{fig:num_arc_disj_paths} shows that, although the most common scenario is to have a single path between any two ASs ($adp=1$), there are still 1193 occurrences of more than one arc-disjoint path between any two ASs ($adp>1$). This means that there is a large number of extra paths that could be used as backup or to shape the flow of information in this network, if a routing protocol capable of exploiting these paths were to be employed and even after the application of local policies has been considered. Therefore, this numerical analysis demonstrates that the Internet possesses unexploited path diversity which could be used to increase its resilience to failures.

It is necessary to remark that the data sources on which these results are based are not entirely accurate. Still, our analysis concurs with results by Oliveira et al.~\cite{Oliveira2008}, which demonstrated that the Internet has greater connectivity than has been previously reported. Also, lack of routing data for ASs outside Europe caused that many other highly connected ASs could not be included in this study. We expect that, if these ASs would had been included and more reliable data were available, a larger and better connected network than what has been observed here, could have been obtained.

We anticipate that the next generation of inter-AS routing protocols should be able to employ the Internet's unexploited path diversity. This is the main objective of a separate publication in which a new routing framework, which employs complete orders as its appropriate topological unit, is introduced \cite{Arjona-Villicana2009a}. The aim of this new work is to propose a new class of routing protocols which employ such topological blocks to guarantee desirable properties, such as lack of instabilities, by construction.

\bibliographystyle{IEEEtran}
\bibliography{IEEEabrv,PD_references}

\end{document}